\PassOptionsToPackage{table}{xcolor}
\documentclass[prd,showpacs,superscriptaddress,nofootinbib,floatfix,11pt]{revtex4-2}

\usepackage[utf8]{inputenc}
\usepackage[T1]{fontenc}
\usepackage{textgreek}
\usepackage{amsmath,amssymb}
\usepackage{slashed}
\usepackage{subfigure}
\usepackage[colorlinks=true,
breaklinks=true,
urlcolor=magenta,
citecolor=blue]{hyperref}
\usepackage[usenames,dvipsnames]{color}
\usepackage{appendix}
\usepackage{braket,bm}
\usepackage{multirow}
\usepackage{enumitem}
\usepackage{slashed}
\usepackage{orcidlink}
\usepackage{graphicx}
\usepackage{tikz}
\usepackage{tikz-feynman}
\usepackage{footnotebackref}
\usepackage{booktabs}
\usepackage{array}
\usepackage{overpic}
\usepackage{float}
\usepackage{threeparttable}
\usepackage[table]{xcolor}

\allowdisplaybreaks[4]

\newcommand{\itp}{\affiliation{Institute of Theoretical Physics, Chinese Academy of Sciences, Beijing 100190, China}}

\newcommand{\ucas}{\affiliation{School of Physical Sciences, University of Chinese Academy of Sciences, Beijing 100049, China}}

\newcommand{\uestc}{\affiliation{School of Physics, University of Electronic Science and Technology of China, Chengdu 611731, China}}

\newcommand{\scnt}{\affiliation{Southern Center for Nuclear-Science Theory (SCNT), Institute of Modern Physics,\\ Chinese Academy of Sciences, Huizhou 516000, China}}

\newcommand{\bonn}{\affiliation{Helmholtz-Institut f\"ur Strahlen- und Kernphysik,
Bethe Center for Theoretical Physics
and Cluster of Excellence  ``Color meets Flavor'', Universit\"at Bonn, D-53115 Bonn, Germany}
}

\newcommand{\fzj}{\affiliation{Institute for Advanced Simulation (IAS-4), Forschungszentrum J\"ulich, D-52425 J\"ulich, Germany}}

\newcommand{\peng}{\affiliation{Peng Huanwu Collaborative Center for Research and Education,
International Institute for Interdisciplinary and Frontiers, Beihang University, Beijing 100191, China}}

\begin{document}

\title{Extraction of the pion-nucleon coupling constant using the effective-range expansion with the left-hand cut}

\author{Bo-Yang Liu}\itp\ucas

\author{Bing Wu\orcidlink{0009-0004-8178-3015}}\email{wu.bing@uestc.edu.cn}
\uestc 

\author{Ji-Wei Fu}\uestc

\author{Meng-Lin~Du\orcidlink{0000-0002-7504-3107}}\email{ du.ml@uestc.edu.cn}
\uestc

\author{Feng-Kun Guo\orcidlink{0000-0002-2919-2064}}\email{fkguo@itp.ac.cn}
\itp \ucas \scnt

\author{Ulf-G. Mei{\ss}ner\orcidlink{0000-0003-1254-442X}}\email{meissner@hiskp.uni-bonn.de}
\bonn \fzj\peng

\begin{abstract}
We apply the generalized effective-range expansion of Ref.~\cite{Du:2024snq} (Phys.\ Rev.\ Lett.\ 135, 011903 (2025)), which incorporates the left-hand cut from one-pion exchange, to low-energy neutron--proton scattering in the $^1S_0$ and $^3S_1$ channels. The amplitude zero for the center-of-mass momentum near 0.35~GeV in the $^1S_0$ channel is naturally accommodated within this framework. 
We extract the pole position, scattering length, effective range, shape parameter, and the pseudoscalar pion--nucleon coupling constant $g_{\pi N}^2/(4\pi)$ at different expansion orders. The low-energy parameters are stable and consistent with established values, while $g_{\pi N}^2/(4\pi)$ exhibits larger uncertainties. The extraction of $g_{\pi N}^2/(4\pi)$ is data-driven, relying on the analytic constraints from the left-hand cut and phase-shift data within the one-pion-exchange approximation.
Despite larger uncertainties compared to high-precision extractions, the consistency with established values demonstrates that this framework can probe the left-hand-cut singularity.

\end{abstract}
\maketitle
\newpage

\section{Introduction}
The effective range expansion (ERE)~\cite{Bethe:1949yr,Blatt:1949zz} provides a model-independent parametrization of the low-energy two-body scattering amplitude, based on unitarity and analyticity, and has been widely employed in the analysis of near-threshold phenomena. Its importance has grown significantly in recent decades with the discovery of numerous exotic hadronic states and candidates located close to two-body thresholds~\cite{ParticleDataGroup:2024cfk,Liu:2024uxn,Guo:2017jvc,Chen:2022asf,Meng:2022ozq,Brambilla:2019esw,Baru:2021ldu,Dong:2021bvy,Dong:2021juy}. For two-particle $S$-wave scattering in relativistic kinematics, the unitarized amplitude takes the form
\begin{align}
    T(s)=8\pi\sqrt{s}f(s)=\frac{8\pi\sqrt{s}}{k\cot\delta-ik}~.\label{unitaryofT}
\end{align}
With this convention, the $S$-wave amplitude $f$ has the dimension of length.
Here, $s$ is the  center-of-mass (c.m.) energy squared, $k$ is the magnitude of the momentum of either particle in the c.m. frame, and \( \delta \) is the phase shift. The conventional ERE expands $k\cot\delta$ as a power series in $k^2$ in the near-threshold region:
 \begin{align}
 \label{eq:ere0}
  k \cot \delta = \frac{1}{a} + \frac{1}{2} r k^2 + v_2 k^4 +\mathcal{O}(k^6)~,
 \end{align}
which defines the scattering length $a$, effective range $r$, and shape parameter $v_2$, with the sign of the first term being convention-dependent. The unitarity cut is isolated in the $ik$ term in Eq.~\eqref{unitaryofT} and $k\cot\delta$ is free of singularities in the vicinity of the threshold. The radius of convergence of the Taylor-series expansion in Eq.~\eqref{eq:ere0} is limited by the location of the nearest singularity in the complex energy plane. When the exchanged particle is relatively light for the $t$-channel exchange (or only slightly heavier than the mass difference of the two scattering particles for the $u$-channel exchange), the branch point of the left-hand cut (lhc) can lie close to the threshold, severely restricting the convergence domain of Eq.~\eqref{eq:ere0}~\cite{Du:2024snq,Du:2025vkm}. See also Ref.~\cite{Epelbaum:2012vx} for a pedagogical introduction.

To systematically incorporate such a near-threshold lhc, a generalized ERE was recently proposed in Ref.~\cite{Du:2024snq},\footnote{The reliability and a higher-order improvement of this scheme, along with a kinematically relativistic version, have been addressed in Ref.~\cite{Wang:2026ups}.} given by
 \begin{align}
k\cot{\delta}=\frac{\tilde{d}(k^2)- \tilde{g} d^\text{R}(k^2)}{\tilde{n}(k^2) + \tilde{g}\left[ L(k^2)-L(0)\right]}\ ,\label{eq:modfere}
 \end{align}
where $\tilde{d}(k^2)$ and the normalized $\tilde{n}(k^2)$ are polynomials in $k^2$. The logarithmic functions $d^\text{R}(k^2)$ and $L(k^2)$, which are given below, encapsulate the analytic structure introduced by the lhc, while the parameter $\tilde{g}$ quantifies its contribution and encodes the coupling strength to the exchanged particle. 
Other approaches that explicitly incorporate the long-range potential arising from the one-particle exchange contributions include the modified L\"uscher formalism~\cite{Raposo:2023oru,Bubna:2024izx,Bubna:2025gsd,Raposo:2025dkb} and effective-field-theory constructions using a plane-wave basis~\cite{Meng:2021uhz,Meng:2023bmz}, which adapt the finite-volume quantization condition to the long-range one-pion-exchange (OPE) force in order to extract infinite-volume scattering amplitudes from lattice QCD energy levels. A different and earlier strategy in Ref.~\cite{Steele:1998zc} employs a modified ERE to remove the long-range (pion) physics from the short-range coefficients; it requires as input the phase shift and Jost function of the specific long-range potential, which for realistic interactions can typically only be obtained numerically. 
Closely related is Ref.~\cite{Baru:2015ira}, which exploits the same long-range OPE to predict correlations among the effective-range-expansion coefficients directly in the infinite volume, but it can also be used to analyze lattice
data, see e.g.~\cite{Baru:2016evv,Lahde:2019yvr,Meyer:2024auq}.
In contrast, our generalized ERE~\eqref{eq:modfere} yields an analytic expression for $k\cot\delta$ that can be directly fitted to phase-shift data, making it particularly suitable for analyzing existing experimental data or serving as a consistency check for lattice QCD extrapolations.

\begin{figure}[tb]
  \centering  
  \includegraphics[width=0.8\textwidth]{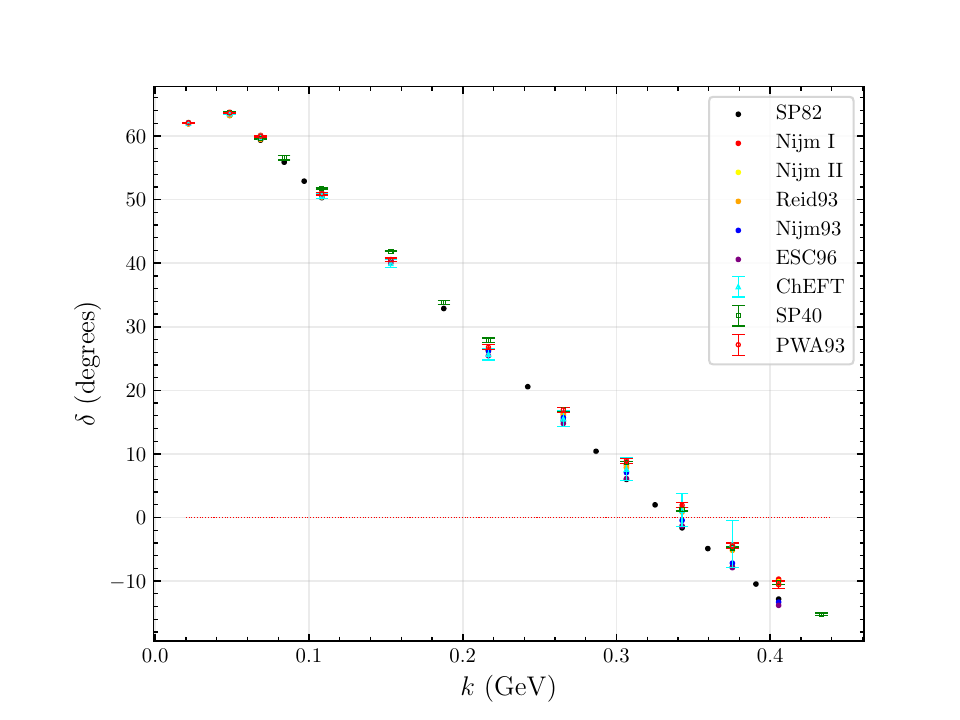}
  \caption{$^1S_0$ $np$ scattering phase shift. Data points are from the following references: SP82 (Ref.~\cite{Arndt:1982ep}), Nijm~I, Nijm~II, Reid93, and Nijm93 (Ref.~\cite{Stoks:1994wp}), ESC96 (Ref.~\cite{Rijken:1995pw}), ChEFT (Ref.~\cite{Reinert:2017usi}), SP40 (Ref.~\cite{Arndt:2000xc}), and PWA93 (Ref.~\cite{Stoks:1993tb}).}
  \label{npscattering data}
\end{figure}

Since the discontinuity of the amplitude across the one-particle exchange lhc originates from the corresponding tree-level exchange diagram, one can extract the coupling strength of the scattering particles to the exchanged particle from the generalized ERE~(\ref{eq:modfere}), as demonstrated for the case of $D^*D\pi$ coupling at an unphysical pion mass in Ref.~\cite{Du:2024snq}. Furthermore, under certain conditions, the generalized ERE can naturally accommodate an amplitude zero, corresponding to a pole of $k\cot\delta$. Since its location can be expressed in terms of ERE parameters and the mass and coupling of the long-range exchanged particle, the corresponding amplitude zero can be understood as arising from an interplay between the short-range and long-range forces~\cite{Du:2024snq}, as seen in Ref.~\cite{Zhang:2023wdz} for neutron--one-neutron-halo-nucleus scattering and in Ref.~\cite{Du:2023hlu} for $DD^*$ scattering. This feature is inherent to the framework. In contrast, the conventional ERE~(\ref{eq:ere0}) lacks such zeros unless augmented by additional Castillejo-Dalitz-Dyson (CDD) poles; however, the connection to the long-range force is missing. Such a near-threshold amplitude zero potentially offers a promising testbed for validating this scheme. Over the past several decades, experimental and theoretical studies of the $^1S_0$ neutron-proton ($np$) phase shift have consistently indicated the presence of an amplitude zero (corresponding to the point $\delta=0$) near $k\simeq 0.35$~GeV, as illustrated in Fig.~\ref{npscattering data}.\footnote{In fact, systematic uncertainties arising from potential model dependence in the extraction of scattering phase shifts from low-energy nucleon-nucleon interaction datasets are typically an order of magnitude larger than statistical uncertainties~\cite{NavarroPerez:2014ovp}. Moreover, when these substantial systematic errors are accounted for, partial-wave correlations in the phase shifts effectively vanish~\cite{Escalante:2020nbq}.} 
Given the presence of both OPE as a long- but finite-range interaction and shorter-range forces, the nucleon--nucleon ($NN$) scattering is an ideal case to apply and test the generalized ERE of Eq.~(\ref{eq:modfere}). Once the OPE lhc is subtracted explicitly, the convergence of the generalized ERE is controlled by the remaining singularities. The nearest one is the two-pion-exchange (TPE) lhc, with branch point at $k=iM_\pi$, which formally sets the radius of convergence of the residual low-energy expansion. {The TPE contribution is, however, of higher chiral order (next-to-leading order), and its lhc branch point lies farther from the physical region than that of the OPE. 
As shown in Ref.~\cite{Rentmeester:1999vw}, adding TPE allows for a reduction in the number of short-range parameters, while $\chi_{\rm min}^2$ decreases only slightly (see Table II therein). This indicates that, although TPE is important for a precise description of $NN$ amplitudes, an OPE-only parametrization with additional short-range parameters can effectively absorb part of the TPE contributions for small $|k^2|<M_\pi^2$, which however could lead to a difference of the coupling determined here compared to the values in the literature.
} 
Beyond this formal radius the generalized ERE~\eqref{equ:kcotdeltawithlhc} is better regarded as a Pad\'e-like parametrization constrained by the explicitly subtracted OPE cut, rather than as a strictly convergent Taylor series. On the physical axis, the next relevant structure is the $NN\pi$ inelastic threshold at $k\approx 0.37$~GeV ($T_{\rm lab}\approx 0.29$~GeV), above which there are inelastic corrections that remain numerically small over the energy range fitted here. 
The $^1S_0$ amplitude zero at $k\approx 0.35$~GeV lies just below this threshold, within the region where the parametrization is reliable.

This framework has been successfully applied to reproduce the \( D D^* \) scattering amplitudes obtained from the Lippmann-Schwinger equation (LSE) and to extract the axial-vector $D^*D\pi$ coupling constant~\cite{Du:2024snq}. It should be noted that in that work, the fit to the LSE amplitude employed six equidistant pseudo-data points, three of which lie on the lhc, including one very close to the lhc branch point. However, since experimental data are confined to the physical region above the threshold, critical questions arise: If only such physical-region data are available, how reliable is this parametrization? To what extent can the coupling constants be accurately extracted?\par
In this work, we analyze low-energy $NN$ scattering phase shifts in the physical region using the generalized ERE, Eq.~(\ref{eq:modfere}), with particular attention to the $^1S_0$ $np$ channel. We extract the pseudoscalar $\pi NN$ coupling constant, as well as the pole position, scattering length, and effective range. We compare the results across different expansion orders with established determinations in the literature to assess the stability of the extraction. The paper is organized as follows. After this Introduction, we present the theoretical formalism in Sec.~\ref{formalism}. This includes a detailed derivation of the generalized ERE for completeness in Sec.~\ref{Chapter IIA ERE with lhc}, the construction of the $NN$ scattering potential in Sec.~\ref{Chapter IIB NNpotential}, and the definitions of physical parameters in Sec.~\ref{parametersderivation}. The fitting of experimental $NN$ data with the generalized ERE at different orders is discussed in Sec.~\ref{analysis of NN data}. The paper concludes with a brief summary in Sec.~\ref{summarychapter}.

\section{Formalism}\label{formalism}
\subsection{ERE with the lhc}\label{Chapter IIA ERE with lhc}
For $S$-wave elastic scattering, the contribution from the $t$-channel one-particle exchange lhc is given by~\cite{Du:2024snq}
\begin{align}
  &L(k^2)\equiv\frac{1}{2}\int_{-1}^{+1}\frac{\mathrm{d}\cos \theta}{t-m_{\rm ex}^2}=-\frac{1}{4k^2}\log\frac{m_{\rm ex}^2/4+k^2}{m_{\rm ex}^2/4} ,\label{equ: left-hand cut}
\end{align}
where $m_{\rm ex}$ denotes the mass of the exchanged particle.\footnote{In this work, isospin-breaking effects between protons and neutrons are neglected, rendering the partial-wave amplitudes of the $u$- and $t$-channels identical.} The branch point of the lhc is located at $k_{\rm lhc}^2=-m_{\rm ex}^2/4$. This implies that a larger exchanged-particle mass places the branch point farther from threshold, thereby reducing the influence of this singularity on the physical region within the dispersion-relation framework. Therefore, in the subsequent analysis, we restrict ourselves to OPE and neglect contributions from heavier exchanges such as $S$-wave two pions (or $\sigma$ in the one-boson exchange model), $\rho$, and $\omega$ mesons.

To simultaneously incorporate both the OPE lhc and the unitarity cut, the formalism of Ref.~\cite{Du:2024snq} starts with the $N/D$ method~\cite{Chew:1960iv,Oller:2019rej}, in which the scattering amplitude is written as 
\begin{align}
  f(k^2)=\frac{n(k^2)}{d(k^2)},\label{equ: f=n/d}
\end{align}
where the numerator $n(k^2)$ contains only the lhc and the denominator $d(k^2)$ contains only the unitarity cut. Using dispersion relations, these functions can be expressed as~\cite{Du:2024snq}
\begin{align}
  &n(k^2)=\tilde{n}(k^2)+\tilde{g}(L(k^2)-L_0),\label{equ: n} \\
  &d(k^2)=\tilde{d}(k^2)-ikn(k^2)-\tilde{g}d^R(k^2),\label{equ: d}
\end{align}
{where $\tilde n(k^2)$ and $\tilde d(k^2)$ are meromorphic functions introduced in Ref.~\cite{Du:2024snq}}. The parameter $\tilde{g}$, together with $L_0=L(k^2=0)=-1/m_{\rm ex}^2$, is chosen so that $n(k^2)$ is normalized at threshold, i.e., $\tilde{n}(0)=1$.\footnote{{In the general formulation of Refs.~\cite{Du:2024snq,Du:2025vkm,Wang:2026ups}, the residual functions (including $\tilde{g}(k^2)$) are analytic in a specified domain before fixing the normalization. After dividing the numerator and denominator by the same polynomial so that $\tilde g$ becomes a constant and imposing $\tilde n(0)=1$, the redefined $\tilde n(k^2)$ and $\tilde d(k^2)$ are in general meromorphic and may contain poles. Over a finite physical energy interval, however, the smooth energy dependence induced by more distant singularities can be effectively absorbed into the residual polynomials, including possible isolated poles (see, e.g., the last paragraph of Sec.~III A in Ref.~\cite{Wang:2026ups}) and the TPE lhc. In practice, one can first try to approximate $\tilde n(k^2)$ and $\tilde d(k^2)$ directly by finite-degree polynomials for simplicity, i.e., the $[m,n]$ parametrization in Eq.~\eqref{equ: ERE with the left-hand cut}.}} The function $d^R(k^2)$, which cancels the lhc contribution from $-ikn(k^2)$ and ensures that $d(k^2)$ is free of the lhc, is given by
\begin{align}
  d^R(k^2)=\frac{i}{4k}\log\frac{m_{\rm ex}/2+ik}{m_{\rm ex}/2-ik}.
  \label{equ:dR}
\end{align}
Expanding $\tilde{n}(k^2)$ and $\tilde{d}(k^2)$ as polynomials of orders $m$ and $n$, respectively, the scattering amplitude can be parameterized as
\begin{align}
  \frac{1}{f_{[m,n]}(k^2)}=\frac{\sum^n_{i=0}\tilde{d}_i k^{2i}-\tilde{g}d^R(k^2)}{1+\sum^m_{j=1}\tilde{n}_j k^{2j}+\tilde{g}(L(k^2)-L_0)}-ik,\label{equ: ERE with the left-hand cut}
\end{align}
where $\tilde{d}_i, i=0,...,n$, and $\tilde{n}_j, j=1,...,m$, are the expansion coefficients. Comparing with the parametrization of the scattering amplitude in Eq.~\eqref{unitaryofT}, we obtain \cite{Du:2024snq}:
\begin{equation}
k\cot\delta=\frac{\sum^n_{i=0}\tilde{d}_i k^{2i}-\tilde{g}d^R(k^2)}{1+\sum^m_{j=1}\tilde{n}_j k^{2j}+\tilde{g}(L(k^2)-L_0)}\ .\label{equ:kcotdeltawithlhc}
\end{equation}
\par

\subsection{The $S$-wave $NN$ OPE potential and unitarized amplitude}\label{Chapter IIB NNpotential}

The Lagrangian for the pseudoscalar form of the $\pi N N$ coupling, together with the nucleon kinetic and mass terms, is given by~
\begin{align}
    \mathcal{L}_{\pi NN}=
    \bar{N}(i\gamma^\mu{\partial_\mu}-m_N)N+ 
    ig_{\pi N}\bar{N}\gamma_5\vec{\tau}\cdot\vec{\pi}N,\label{equ:Lag}
\end{align}
where $m_N$ is the isospin-averaged nucleon mass,\footnote{Strictly speaking, the chiral limit values of the various parameters appear here, but to the order we are working, we can equate these with their physical values.} $g_{\pi N}$ is the pseudoscalar pion-nucleon coupling constant, and $\vec \tau$ denotes the Pauli matrices in the isospin space. 
The isospin multiplets are defined as
\begin{align}
    N=\left(\begin{array}{c}
    p\\
    n \end{array}\right),\quad\vec{\pi}=\left( \frac{\pi^++\pi^-}{\sqrt{2}}, \frac{\pi^--\pi^+}{i\sqrt{2}},\pi^0 \right).
\end{align}
Alternatively, the $\pi NN$ coupling can take a derivative form to reflect the fact that pions are pseudo-Nambu-Goldstone bosons of the spontaneous chiral symmetry breaking, for which the standard leading-order SU(2) chiral Lagrangian for the $\pi N$ interaction gives
\begin{align}
\mathcal{L}_{\pi N}^{(1)}= \frac{g_A}{2}\bar{N} 
\gamma^\mu\gamma_5u_\mu  N ,\label{equ:chiralLag}
\end{align}
where $g_A$ denotes the nucleon axial-vector coupling constant in the chiral limit. The chiral vielbein is defined as
\begin{align}
  u_\mu = i u^{\dagger} \partial_\mu u-i u \partial_\mu u^{\dagger}, 
\end{align}
with $u^2= U$ and $U = \exp \left(i \vec \tau \cdot \vec \pi/{F_\pi}\right)$,
where $F_\pi$ is the pion decay constant in the chiral limit. The $\pi NN$ couplings in the Lagrangians~\eqref{equ:Lag} and \eqref{equ:chiralLag} for on-shell particles are equivalent, and the coupling constants are related to each other by the Goldberger-Treiman relation, $g_{\pi N}=g_A m_N /F_\pi$~\cite{Goldberger:1958zz}. Corrections to this relation are small; see, e.g., Ref.~\cite{Fettes:1998ud}.

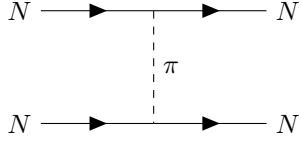
\begin{figure}[t]
    \centering
    \begin{tikzpicture}
        \begin{feynman}
            \vertex (a);
            \vertex [left=of a] (i1) {\(N\)};
            \vertex [right=of a] (f1) {\(N\)};
            \vertex [below=1.5cm of a] (b);
            \vertex [left=of b] (i2) {\(N\)};
            \vertex [right=of b] (f2) {\(N\)};
            \diagram* {
                (i1) -- [fermion] (a) -- [fermion] (f1),
                (i2) -- [fermion] (b) -- [fermion] (f2),
                (a) -- [scalar, edge label=\(\pi\)] (b),
            };
        \end{feynman}
    \end{tikzpicture}
    \caption{Feynman diagram for the $NN\to NN$ process with OPE.}
    \label{fig:one_pion_exchange}
\end{figure}

From the  Lagrangian~\eqref{equ:Lag}, the tree-level OPE amplitude\footnote{In the c.m. frame, the $t$-channel momentum transfer is purely spatial, $q = (0, \vec{q}\,)$, the amplitude reduces to the three-dimensional form given in Eq.~(\ref{equ:NN amplitude}).}  in the c.m. frame as shown in Fig.~\ref{fig:one_pion_exchange} reads
\begin{align}
    V(\vec{q\,})=g_{\pi N}^2 \vec{\tau}_1\cdot\vec{\tau}_2\frac{\vec{\sigma}_1\cdot\vec{q}\ \vec{\sigma}_2\cdot\vec{q}}{\vec{q\,}^2+M_\pi^2} ,\label{equ:NN amplitude}
\end{align}
where $M_\pi$ is the isospin-averaged pion mass, $\vec{\sigma}_i$ ($\vec{\tau}_i$) is the spin (isospin) Pauli matrix of nucleon $i$, and $\vec{q}$ is the three-momentum transferred by the pion. Using the SU(2) Clebsch–Gordan coefficients, the isospin factor evaluates to
\begin{align}
\langle I_f|\vec{\tau}_1\cdot\vec{\tau}_2|I_i\rangle=\left( 2I_f(I_f+1) -3\right)\delta_{I_fI_i}\ ,
\end{align}
where $I_i$ and $I_f$ denote the total isospin of the initial and final two-nucleon states, respectively. For $S$-wave scattering, the spin-dependent term upon angular averaging simplifies as
\begin{align}
  \vec{\sigma}_1\cdot\vec{q}\ \vec{\sigma}_2\cdot\vec{q} \to \frac{\vec{\sigma}_1\cdot\vec{\sigma}_2}{3}\vec{q\,}^2 .
\end{align}
This follows from $\langle q_i q_j\rangle_\Omega=\frac{1}{3}\vec{q\,}^2\delta_{ij}$ for an $S$-wave angular average.
Due to the Pauli exclusion principle, the total (isospin, spin) of an $S$-wave $NN$ system could be either (1,0) or (0,1).  As a consequence, the states $|^1S_0,I=1\rangle$ and $|^3S_1,I=0\rangle$ experience the same potential
\begin{align}
V_0(k^2)=-g_{\pi N}^2M_\pi^2L(k^2)\ ,\label{equ: VOPE}
\end{align}
where the subscript 0 indicates the $S$-wave.
For the $^1S_0$ ($I=1$) and $^3S_1$ ($I=0$) channels, the spin--isospin factor yields the same projected OPE kernel $V(\vec{q\,})=-g_{\pi N}^2\,\vec{q\,}^2/(\vec{q\,}^2+M_\pi^2)$. Writing $\vec{q\,}^2/(\vec{q\,}^2+M_\pi^2)=1-M_\pi^2/(\vec{q\,}^2+M_\pi^2)$, the first term is a short-range contact interaction (omitted here), while the second term gives Eq.~(\ref{equ: VOPE}) after the $S$-wave projection with $t=-\vec{q\,}^2$.
Note that $L(0)\neq 0$; however, in the $N/D$ representation the lhc contribution enters through the subtracted combination $L(k^2)-L_0$ in Eq.~(\ref{equ: n}), which fixes the threshold normalization and does not affect the lhc discontinuity determined by ${\rm Im}\,L(k^2)$.
Starting from this tree-level OPE and contact potentials, the unitarized amplitude $T(k^2)$ can be obtained by solving the LSE.
Noting that the discontinuity of the amplitude across the one-particle exchange lhc originates from the corresponding tree-level exchange diagram, we obtain~\cite{Du:2024snq}
\begin{align}
\mathrm{Im}T(k^2)=8\pi\sqrt{s}\,\mathrm{Im}f(k^2)=\mathrm{Im}V_0(k^2),\quad \text{for }  k^2\leq k^2_{\rm lhc}. \label{equ:ImT}
\end{align}

\subsection{Extraction of low-energy parameters}\label{parametersderivation}
Now we can relate the generalized ERE with lhc and OPE potential to extract the coupling constant following Ref.~\cite{Du:2024snq}. From Eqs.~(\ref{equ: f=n/d}, \ref{equ: n}) and noting that $d(k^2)$ is free of the lhc, we obtain
\begin{align}
  \mathrm{Im}f(k^2)=\frac{\tilde{g}\,{\rm Im}L(k^2)}{d(k^2)},\quad \text{for }k^2\leq k^2_{\rm lhc}.
\end{align} 
Combining Eqs.~(\ref{equ: VOPE}, \ref{equ:ImT}) and restricting our analysis to the near-threshold region, we obtain  
\begin{align}
    \frac{g^2_{\pi N}}{4\pi}=-\frac{\tilde{g}}{d^{0,{\rm lhc}}}\frac{4m_N}{M_\pi^2},
\end{align}
where $d^{0,{\rm lhc}}$ denotes the value of $d(k^2)$ at the branch point of the lhc, i.e.,
\begin{align}
  &d^{0,{\rm lhc}}=\sum^n_{i=0}\tilde{d}_i\left(\frac{-m^2_{\rm ex}}{4}\right)^i+\frac{m_{\rm ex}}{2}\left(1+ \sum_{j=1}^m\tilde{n}_j\left(\frac{-m^2_{\rm ex}}{4}\right)^j +\frac{\tilde{g}}{m^2_{\rm ex}}\right)+\frac{\tilde{g}\log 2}{m_{\rm ex}}.\label{equ: dlhc}
\end{align}
This expression is obtained by evaluating Eq.~(\ref{equ: d}) in the limit $k^2\to -m_{\rm ex}^2/4$ (i.e., $k\to i m_{\rm ex}/2$), where the logarithmic singularities in $-ikn(k^2)$ and $d^R(k^2)$ cancel, leaving the finite $\log 2$ term; see Ref.~\cite{Du:2024snq} for details.

Furthermore, from Eqs.~(\ref{eq:ere0}, \ref{equ:kcotdeltawithlhc}), the $S$-wave scattering length $a$, effective range $r$ and shape parameter $v_2$ can be extracted directly as
\begin{align}\label{eq:arv2}
  a&=f(k^2=0)=\left(\tilde{d}_0+\frac{\tilde{g}}{m_{\rm ex}}\right)^{-1},\notag\\
  r&=2\left.\frac{\mathrm{d}}{\mathrm{d}(k^2)}\left(\frac{1} {f}+ik\right)\right|_{k^2=0}=2\tilde{d}_1-\frac{8\tilde{g}}{3m_{\rm ex}^3}-\left(2\tilde{n}_1+\frac{4\tilde{g}}{m_{\rm ex}^4}\right)\frac{1}{a},\notag\\
  v_2 &=\frac{1}{2}\left.\frac{\mathrm{d}^2}{\mathrm{d}(k^2)^2}\left(\frac{1} {f}+ik\right)\right|_{k^2=0}=\tilde{d}_2 + \frac{16}{5}\frac{\tilde{g}}{m_{\text{ex}}^5} 
  - \left(\tilde{n}_1 + \frac{2\tilde{g}}{m_{\text{ex}}^4}\right) \left(\tilde{d}_1 - \frac{4\tilde{g}}{3m_{\text{ex}}^3}\right) \notag\\
  &\qquad\qquad\qquad\qquad\qquad\qquad\quad+ \left[ \left(\tilde{n}_1 + \frac{2\tilde{g}}{m_{\text{ex}}^4}\right)^2 - \tilde{n}_2 + \frac{16\tilde{g}}{3m_{\text{ex}}^6} \right] \frac{1}{a}.
\end{align}
The coefficients in the effective-range expression follow from the threshold expansions of Eqs.~(\ref{equ: left-hand cut}, \ref{equ:dR}):
\begin{align*}
  L(k^2)= -\frac{1}{m_{\rm ex}^2}+\frac{2k^2}{m_{\rm ex}^4}-\frac{16k^4}{3m_{\rm ex}^6}+\mathcal{O}(k^6),\qquad d^R(k^2)= -\frac{1}{m_{\rm ex}}+\frac{4k^2}{3m_{\rm ex}^3}-\frac{16k^4}{5m_{\rm ex}^5}+\mathcal{O}(k^6).
\end{align*}
Expanding Eq.~(\ref{equ:kcotdeltawithlhc}) to $\mathcal{O}(k^4)$ and using $k\cot\delta=1/f+ik$ then reproduces the numerical coefficients given in Eq.~(\ref{eq:arv2}).

\section{Results and discussion}\label{analysis of NN data}

Using the $np$ scattering phase shifts from Ref.~\cite{Stoks:1993tb},\footnote{We use $np$ rather than proton-proton ($pp$) scattering data because the latter involves additional Coulomb effects~\cite{Wiringa:1994wb} that fall outside the present theoretical framework.} which provide 11 data points for each of the $|^1S_0,I=1\rangle$ and $|^3S_1,I=0\rangle$ channels below 350 MeV for the kinetic energy in the laboratory frame (see Fig.~\ref{Fig.main}) and serve as a benchmark in chiral nuclear force studies (see, e.g., Refs.~\cite{Epelbaum:2014sza,Epelbaum:2014efa,Lu:2021gsb}), we perform fits with the generalized ERE as given in Eq.~\eqref{equ:kcotdeltawithlhc}.

As mentioned above, it allows us to extract the low-energy parameters for the $NN$ system, including the pseudoscalar coupling constant, pole position, scattering length, and effective range, and it also serves to test the stability of the present parametrization scheme. Two different fitting strategies are employed. In the first strategy, we fit the phase shifts of the independent channels $|^1S_0,I=1\rangle$ and $|^3S_1,I=0\rangle$ (the deuteron channel) separately.\footnote{The effects of the $S$-$D$ mixing due to the tensor force, via the term $\vec{\sigma}_1\cdot\vec{q}\ \vec{\sigma}_2\cdot\vec{q}$ in Eq.~(\ref{equ:NN amplitude}), are neglected in this analysis because the $D$-state probability in the deuteron is only about 5\%~\cite{Machleidt:1987hj} and does not affect the location of the lhc branch point. Accordingly, the $|^3S_1\rangle$ phase shift is fitted as a single (uncoupled) $S$ wave, and the extracted $r$ and $v_2$ are compared in Tables~\ref{tab:1S03S1fit} and~\ref{tab:v2fit} with the corresponding single-channel determinations. As the tensor force is the dominant long-range interaction in this channel, the extracted $|^3S_1\rangle$ coupling is less controlled than that from $|^1S_0\rangle$ and is regarded as illustrative.} The second strategy consists in performing a combined analysis by fitting to both channels simultaneously while incorporating the high-precision experimental measurement of the bound coherent scattering length
\begin{align}\label{eq:coherent}
    a_{\rm c}=\left(1+\frac{m_n}{m_p}\right)\frac{1}{4}\left(3a_{\ket{^3S_1}}+a_{\ket{^1S_0}}\right),
\end{align}
where $m_n$ and $m_p$ denote the neutron and proton masses, respectively. The prefactor $(1+m_n/m_p)$ accounts for the standard conversion between the c.m. and laboratory definitions of the bound coherent scattering length. In the isospin-symmetric approximation $m_n=m_p$ adopted in this work, Eq.~(\ref{eq:coherent}) reduces to $a_{\rm c}=\frac{1}{2}(3a_{\ket{^3S_1}}+a_{\ket{^1S_0}})$. A quantitative estimate of isospin-breaking corrections to this constraint is beyond the scope of the present work.
We take $a_{\rm c}=3.7384(20)\,\mathrm{fm}$ from Ref.~\cite{Schoen:2003my},\footnote{The sign of the coherent scattering length quoted in Ref.~\cite{Schoen:2003my} has been converted to match the convention adopted in this work.} which was measured via neutron interferometry in gaseous molecular hydrogen, entirely independent of the phase shift data used in our fit.

\newpage
\subsection{Channel-by-channel Fit}

\begin{figure}[th]
  \centering  
  \includegraphics[width=0.7\textwidth]{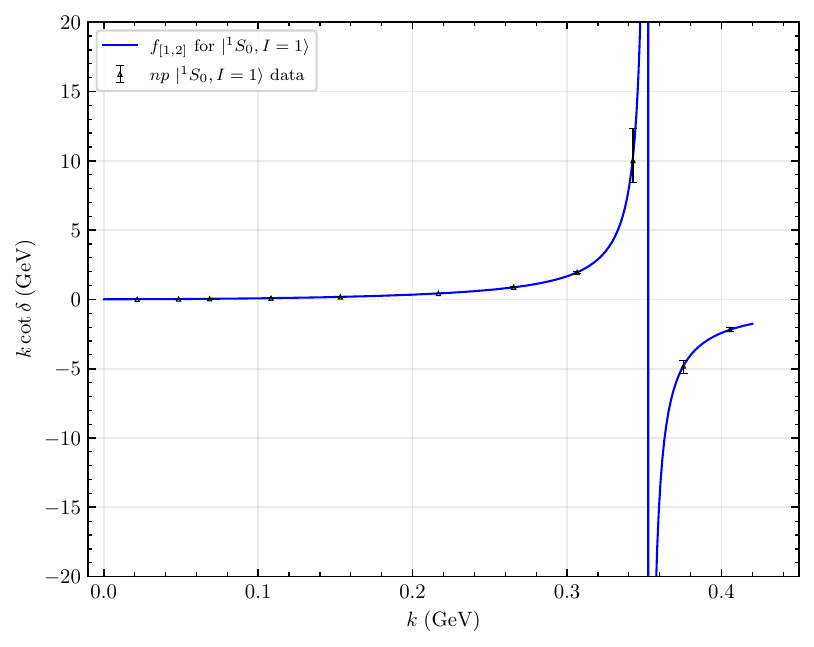}
  \caption{Comparison of $k \cot \delta$ between the best fit and data for the $np$ $|^1S_0,I=1\rangle$ channel (fit performed with the generalized ERE~\eqref{equ:kcotdeltawithlhc} on scattering phase shift from Ref.~\cite{Stoks:1993tb}).}
  \label{Fig.mainkcot1S0}
\end{figure}

An excellent description of the $|^1S_0,I=1\rangle$ channel phase shift is achieved with the $[1,2]$ order parametric scheme\footnote{An almost equally good description of the phase shift can be achieved with the $[2,1]$ order parametrization, leading to consistent low-energy parameters and the pole position with the $[1,2]$ parametrization.} in Eq.~(\ref{equ: ERE with the left-hand cut}). The extracted parameters, a virtual pole at $-66.4^{+0.6}_{-0.5}$ keV on the unphysical Riemann sheet, a scattering length of $23.72^{+0.10}_{-0.08}$ fm, and an effective range of $2.68(3)$ fm (row 5 of Table~\ref{tab:1S03S1fit}), show excellent agreement with earlier results from Ref.~\cite{Babenko:2007ss,Dumbrajs:1983jd,Machleidt:2000ge} (rows 9 to 11), where the pole positions in Refs.~\cite{Babenko:2007ss,Dumbrajs:1983jd,Machleidt:2000ge} are derived from the conventional ERE~\cite{Epelbaum:2008ga} $E_{\rm pole} = (1-\sqrt{1+2r/a})^2/(r^2 m_N)$.
The pseudoscalar coupling constant, ${g_{\pi N}^2}/(4\pi)=12.4_{-2.3}^{+2.2}$, also falls within the accepted range and is compatible with previous determinations~\cite{Reinert:2020mcu,Arndt:2006bf,Baru:2011bw,Arriola:2025dnh,Timmermans:1990tz} (rows 14 to 18) within the conditional uncertainties discussed above. In particular, the amplitude zero, corresponding to a pole of $k\cot\delta$, at around $k=0.35$ GeV, see Fig.~\ref{Fig.mainkcot1S0}, is properly produced.

For the $|^3S_1,I=0\rangle$ channel, the $[2,1]$ parametric scheme provides a good description of the data, yielding a bound pole at $-2222^{+14}_{-12}$ keV, a scattering length of $-5.416^{+0.002}_{-0.003}$ fm, and an effective range of $1.749(7)$ fm (row 7 of Table~\ref{tab:1S03S1fit}), in good agreement with earlier studies~\cite{Babenko:2007ss,Stoks:1993tb,NavarroPerez:2013mvd,Dumbrajs:1983jd,Machleidt:2000ge} (rows 9 to 13). In contrast, the pseudoscalar coupling constant exhibits a larger uncertainty, specifically $g_{\pi N}^2/(4\pi) = 18.3_{-8.6}^{+8.4}$. Nevertheless, within uncertainties, our extracted coupling remains consistent with previous determinations~\cite{Reinert:2020mcu,Arndt:2006bf,Baru:2011bw,Arriola:2025dnh,Timmermans:1990tz}.
The large uncertainty shows that without the $S$-$D$ mixing that arises from the OPE tensor force, the sensitivity from only the $S$-wave phase shift to the OPE is weaker than that in the $|^1S_0,I=1\rangle$ channel.\par

The extraction of the coupling constant $g_{\pi N}^2/(4\pi)$ provides a nontrivial consistency check of the OPE-driven lhc contribution within the generalized ERE framework, and shows that the framework can be used to probe couplings related to the one-particle exchange. In contrast to the low-energy parameters $(a,r,E_{\rm pole})$, which are tightly constrained by the phase shifts, the uncertainty of $g_{\pi N}^2/(4\pi)$ is considerably larger, especially in the $^3S_1$-only fit, reflecting that the sensitivity to the lhc is weaker since it is located in the unphysical region and is logarithmic.

\subsection{Joint analysis of the \texorpdfstring{$^1S_0$}{1S0} and \texorpdfstring{$^3S_1$}{3S1} channels}

In this subsection, we perform a joint analysis of both $|^1S_0,I=1\rangle$ and $|^3S_1,I=0\rangle$ channels under the constraint of the coherent scattering length $a_c=3.7384(20)$ fm, defined in Eq.~\eqref{eq:coherent}. While the expansion orders and the parameter values in Eq.~\eqref{equ: ERE with the left-hand cut} are different in general for the two channels, their long-range potentials are identical, as given in Eq.~\eqref{equ: VOPE}. This common long-range part implies that $\tilde{g}/d^{0,\text{lhc}}$ should be the same for the two channels, leading to the common $g^2_{\pi N}/(4\pi)$. Imposing this constraint together with the central value of the coherent scattering length, we fit the phase shifts of both channels using either the same or different expansion orders. 

\begin{figure}[t]
  \centering 
  \includegraphics[width=0.7\textwidth]{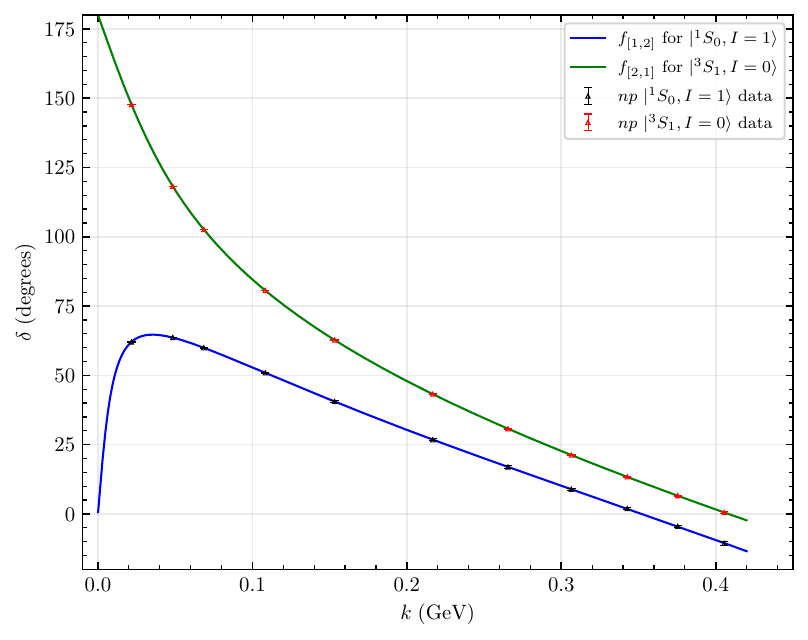}
  \caption{Joint fit to the data from Ref.~\cite{Stoks:1993tb} using the generalized ERE in Eq.~\eqref{equ:kcotdeltawithlhc}.}
  \label{Fig.main}
\end{figure}

For comparison, four distinct parametrizations of Eq.~\eqref{equ: ERE with the left-hand cut} at different orders of the parametrization that provide a good description of the data are considered. 
As shown in the first four rows of Table~\ref{tab:1S03S1fit}, all four sets yield highly consistent values for the pole positions, scattering lengths, effective ranges, and coupling constants. 
Within uncertainties, these results agree with previous determinations (last five rows of Table~\ref{tab:1S03S1fit}).
The extracted central values are stable across these parametrizations; however, the central value of the coupling constant tends to be smaller than several recent high-precision determinations. 
The corresponding fit results are displayed in Fig.~\ref{Fig.main}. Only one representative parametrization is plotted, as the differences among the various schemes are visually indistinguishable. We have also varied the input coherent scattering length within its quoted uncertainty, $a_{\rm c}=3.7384(20)$ fm~\cite{Schoen:2003my}. The resulting uncertainty propagated to $g_{\pi N}^2/(4\pi)$ was found to be extremely small (about 0.07), negligible compared to the statistical errors.

\renewcommand{\arraystretch}{1}
\begin{table}[h]
\caption{Extracted parameters for the $NN$ $|^1S_0,I=1\rangle$ and $|^3S_1,I=0\rangle$ systems from a fit to the $np$ scattering phase shifts of Ref.~\cite{Stoks:1993tb} using the ERE with lhc~(\ref{equ:kcotdeltawithlhc}). The table lists the pole position, scattering length, effective range, pseudoscalar coupling constant, and $\chi^2$/d.o.f. The first four rows present a joint analysis of both channels with a common $g^2_{\pi N}/(4\pi)$ and the fixed bound coherent scattering length $a_{\rm c} = 3.7384$~fm, taken from the central value reported in Ref.~\cite{Schoen:2003my}. Results of independent fits to the $|^1S_0,I=1\rangle$ and $|^3S_1,I=0\rangle$ channels are shown separately in rows 5--8, respectively. For comparison, rows 9--18 list values obtained from other references. The sign of the scattering length quoted in Refs.~\cite{Dumbrajs:1983jd,Machleidt:2000ge} has been converted to match the convention adopted in this work, i.e., $k\cot\delta=1/a+\cdots$ as in Eq.~(\ref{eq:ere0}). The quoted pole positions refer to the first Riemann sheet for the $|^3S_1,I=0\rangle$ state (the deuteron) and the second Riemann sheet for the $|^1S_0,I=1\rangle$ state. Negative values of $E_{\rm pole}$ correspond to poles below the $NN$ threshold.}
\centering
\footnotesize
\begin{tabular}{>{\centering\arraybackslash}p{3em} >{\centering\arraybackslash}p{3em}| >{\centering\arraybackslash}p{5em} >{\centering\arraybackslash}p{6em}| >{\centering\arraybackslash}p{5em} >{\centering\arraybackslash}p{6em}| >{\centering\arraybackslash}p{4.5em} >{\centering\arraybackslash}p{5em}| >{\centering\arraybackslash}p{4.5em} >{\centering\arraybackslash}p{4em}}
\toprule[2pt]
\multicolumn{2}{c|}{ERE w/ lhc} &  \multicolumn{2}{c|}{$E_{\rm pole}$ (keV)} & \multicolumn{2}{c|}{$a$ (fm)} & \multicolumn{2}{c|}{$r$ (fm)} &\multirow{2}*{$\frac{g_{\pi N}^2}{4\pi}$} & \multirow{2}*{$\frac{\chi^2}{\rm d.o.f}$} \\
$|^1S_0\rangle$ & $|^3S_1\rangle$  &$|^1S_0\rangle$ & $|^3S_1\rangle$& $|^1S_0\rangle$& $|^3S_1\rangle$& $|^1S_0\rangle$& $|^3S_1\rangle$& & \\
\midrule[0.5pt]
$f_{[2,1]}$&$f_{[1,2]}$ &$-66.33(4)$ & $-2208(4)$&$23.73(1)$ &$-5.418(2)$ &$2.68(1)$ & $1.741(3)$& $12.2(1.4)$& $\frac{2.58}{22-8}$\\
$f_{[1,2]}$ &$f_{[1,2]}$ &$-66.33(4)$  & $-2210_{-5}^{+4}$ & $23.73(1)$ &$-5.418(2)$ & $2.68(1)$ & $1.742(3)$ & $12.8(1.2)$ & $\frac{1.77}{22-8}$ \\
$f_{[2,1]}$ & $f_{[2,1]}$&$-66.34(4)$ & $-2211(5)$&23.73(1) &$-5.418(2)$ &2.68(1) &1.743(3) & $11.9_{-1.5}^{+1.4}$& $\frac{1.41}{22-8}$\\
$f_{[1,2]}$ & $f_{[2,1]}$ & $-66.34(4)$&$-2212(4)$ & $23.73(1)$ & $-5.418(2)$ & 2.68(1) & 1.744(3) & $12.7(1.2)$& $\frac{0.74}{22-8}$\\
$f_{[1,2]}$& - &$-66.4^{+0.6}_{-0.5}$ & - &$23.72^{+0.10}_{-0.08}$ &- & $2.68(3)$ &- &$12.4_{-2.3}^{+2.2}$ & $\frac{0.05}{11-5}$\\
$f_{[2,1]}$ & - &$-66.6^{+0.6}_{-0.5}$ & - &$23.70^{+0.10}_{-0.08}$ &- & $2.67(3)$ &- &$11.0_{-2.8}^{+2.6}$ & $\frac{0.49}{11-5}$\\
-& $f_{[2,1]}$ &-& $-2222^{+14}_{-12}$ & - & $-5.416^{+0.002}_{-0.003}$&-& 1.749(7)& $18.3_{-8.6}^{+8.4}$& $\frac{0.23}{11-5}$\\ 
-& $f_{[1,2]}$ &-& $-2227^{+13}_{-12}$ & - & $-5.416^{+0.002}_{-0.003}$&-& $1.752(7)$& $22.6(7.6)$& $\frac{0.023}{11-5}$\\
\multicolumn{2}{c|}{\cite{Babenko:2007ss}} & $-66.5$ & $-2237$ & 23.719 & $-5.403$ & 2.626 & 1.749 &-&-\\
\multicolumn{2}{c|}{\cite{Dumbrajs:1983jd}} & $-66.08(13)$ & $-2223(6)$ & $23.748(10)$ & $-5.424(4)$ & 2.75(5) & 1.759(5) &-&-\\
\multicolumn{2}{c|}{\cite{Machleidt:2000ge}} & $-66.08(16)$ & $-2223(10)$ & $23.740(20)$ & $-5.419(7)$ & 2.77(5) & 1.753(8) &-&-\\
\multicolumn{2}{c|}{\cite{Stoks:1993tb} } &-& $-2224.7(35)$&-&-&-&-&-&-\\ 
\multicolumn{2}{c|}{\cite{NavarroPerez:2013mvd} } &-& $-2224.575(9)$&23.74(2)&-&-&-&-&-\\
\multicolumn{2}{c|}{\cite{Arndt:2006bf}} & - & - & - & - & - & - &$13.76(1)$ &- \\ 
\multicolumn{2}{c|}{\cite{Baru:2011bw}} & - & - & - & - & - & - &$13.69(19)$ &- \\ 
\multicolumn{2}{c|}{\cite{Reinert:2020mcu}} & - & - & - & - & - & - &$13.92(9)$ &- \\
\multicolumn{2}{c|}{\cite{Arriola:2025dnh}} & - & - & - & - & - & - &$13.74_{-0.08}^{+0.15}$ &- \\
\multicolumn{2}{c|}{\cite{Timmermans:1990tz}} & - & - & - & - & - & - &$13.6(3)$ &- \\
\bottomrule[2pt]
\end{tabular}
\label{tab:1S03S1fit}
\end{table}

Table~\ref{tab:1S03S1fit} also lists selected results for the pseudoscalar $\pi N$ coupling from earlier studies. Our extracted values of $g_{\pi N}^2/(4\pi)$ fall into the broad range of existing determinations. Following Yukawa's theory, Bethe obtained an early estimate of $g_{\pi N}^2/(4\pi)\approx 14$ in the 1940s by solving the Schrödinger equation for deuteron observables~\cite{Bethe:1940iba,Bethe:1940zz}. Nambu later estimated a value around 13.5 using the Goldberger-Treiman relation~\cite{Nambu:1960xd,Goldberger:1958tr}. 
Through pion pole extrapolation of the renormalized Born approximation~\cite{Chew:1958zz}, analyses of $np$ and $pp$ scattering data yielded $g_{\pi N}^2/(4\pi)=12.0(21)$~\cite{Cziffra:1959zza,MacGregor:1959zz}. Subsequent partial-wave analyses of $NN$ observables further confirmed the OPE contribution, reported values around $14$~\cite{Signell:1960zz} and $12$~\cite{PhysRev.135.B628}, respectively. In the 1990s, multi-energy partial-wave analyses using the Nijmegen potential on the data of Ref.~\cite{Stoks:1993tb} produced $g_{\pi N}^2/(4\pi)=13.56(7)^*$\footnote{Values marked with an asterisk are originally given in terms of $f_{\pi^aNN'}^2$, which relates to the pseudoscalar coupling via ${g_{\pi^aNN'}^2}/(4\pi)=f_{\pi^aNN'}^2(m_{N'} +m_{N})^2/M_{\pi^\pm}^2$ (with $N,N'=n,p$ and $\pi^a=\pi^0,\pi^\pm$). As the relevant masses are not explicitly specified in the cited references, we convert these values using standard particle masses.}~\cite{Klomp:1991vz} and $13.54(5)^*$~\cite{Stoks:1992ja}.
A later study that replaced the Nijmegen potential with a chiral TPE potential obtained $13.68(7)^*$~\cite{Rentmeester:1999vw}. Determinations have also been made from the $\pi N$ data via dispersion relations~\cite{Arndt:2006bf} and the Goldberger-Miyazawa-Oehme sum rule~\cite{Arndt:2006bf,Baru:2011bw}, as well as from $N\bar{N}$ data, which yielded $13.6(3)$~\cite{Timmermans:1990tz}. 
More recent analyses include a covariance-error partial-wave fit to a selected 3$\sigma$-self consistent database, giving $g_{\pi N}^2/(4\pi)=13.98(8)$~\cite{NavarroPerez:2016eli}, and a Bayesian chiral-effective-field-theory analysis with fully controlled uncertainties, reporting $13.92(9)$~\cite{Reinert:2020mcu}.
A theoretical extraction based on pseudoscalar-meson dominance and dispersion relations gives $13.74_{-0.08}^{+0.15}$~\cite{Arriola:2025dnh}. Comprehensive reviews of the history of $g_{\pi N}$ determinations can be found in Refs.~\cite{deSwart:1997ep,Sainio:1999ba,Bugg:2004cm,Matsinos:2019kqi}. We note that the central values extracted in the joint analysis are somewhat smaller than several recent high-precision determinations spreading from about 13.6 to 13.9; this difference may be understood as contributions beyond the present OPE lhc approximation (such as two-pion exchange, $S$-$D$ mixing, and $NN\pi$ and $N\Delta$ coupled channels), which are not included here. The extracted values of $g_{\pi N}^2/(4\pi)$ across different expansion orders and schemes exhibit good internal stability. We highlight that the extraction of $g_{\pi N}^2/(4\pi)$ in this work is data-driven within the OPE lhc approximation adopted here, as it relies primarily on the analytic constraints imposed by the lhc and the scattering data. The spread among different expansion orders in Table~\ref{tab:1S03S1fit} provides a rough estimate of the truncation error. Accordingly, we interpret the extracted $g_{\pi N}^2/(4\pi)$ as an effective measure of the OPE lhc strength within the present truncation, rather than as a competitive high-precision determination of the physical coupling.

\subsection{Extraction of shape parameter $v_2$}
\renewcommand{\arraystretch}{1}
\begin{table}[tb]
  {\caption{Extracted shape parameter $v_2$ for the $NN$ $|^1S_0,I=1\rangle$ and $|^3S_1,I=0\rangle$ systems from a fit to the $np$ scattering phase shifts of Ref.~\cite{Stoks:1993tb} using the ERE with lhc~(\ref{equ:kcotdeltawithlhc}). The notation follows that of Table~\ref{tab:1S03S1fit}. In the row of Refs.~\cite{deSwart:1995ui,PavonValderrama:2005ku}, the $|^3S_1,I=0\rangle$ value is taken from Ref.~\cite{deSwart:1995ui} and the $|^1S_0,I=1\rangle$ value from Ref.~\cite{PavonValderrama:2005ku}, as Ref.~\cite{deSwart:1995ui} reports only the deuteron ($^3S_1$) parameters.}
  \centering
  \footnotesize
  \begin{tabular}{>{\centering\arraybackslash}p{4em} >{\centering\arraybackslash}p{4em}| >{\centering\arraybackslash}p{7em} >{\centering\arraybackslash}p{7em}|>{\centering\arraybackslash}p{6em}}
  \toprule[2pt]
  \multicolumn{2}{c|}{ERE w/ lhc} & \multicolumn{2}{c|}{$v_2$ (fm$^3$)} & \multirow{2}*{$\frac{\chi^2}{\rm d.o.f}$}\\
  $|^1S_0\rangle$ & $|^3S_1\rangle$ & $|^1S_0\rangle$ & $|^3S_1\rangle$ &\\
  \midrule[0.5pt]
$f_{[2,1]}$ & $f_{[1,2]}$ &$-0.52(8)$ &$0.09(1)$ & $\frac{2.58}{22-8}$\\
  $f_{[1,2]}$ & $f_{[1,2]}$ & $-0.56(8)$ & $0.08(1)$ & $\frac{1.77}{22-8}$\\
  $f_{[2,1]}$ & $f_{[2,1]}$ &$-0.51_{-0.09}^{+0.08}$& $0.08(1)$& $\frac{1.41}{22-8}$\\
  $f_{[1,2]}$ & $f_{[2,1]}$ & $-0.54(8)$ & $0.08(1)$ & $\frac{0.74}{22-8}$\\
  $f_{[1,2]}$ & -            & $-0.53(12)$ & - & $\frac{0.05}{11-5}$\\
  $f_{[2,1]}$ & -            & $-0.45(12) $ & - & $\frac{0.49}{11-5}$\\
  -            & $f_{[2,1]}$ & - & $0.05(4)$ & $\frac{0.23}{11-5}$ \\
  -            & $f_{[1,2]}$ & - & $0.03_{-0.03}^{+0.04}$ & $\frac{0.023}{11-5}$\\
  \multicolumn{2}{c|}{\cite{NavarroPerez:2014ovp}} & $-0.48$ & $0.045$ & - \\
  \multicolumn{2}{c|}{\cite{deSwart:1995ui,PavonValderrama:2005ku}} & $-0.5$ & $0.04$ & - \\
  \multicolumn{2}{c|}{\cite{Epelbaum:2015sha,Baru:2015ira}} & $[-0.61,-0.55]$ & $0.06$ & - \\
  \multicolumn{2}{c|}{\cite{Thim:2024jdv}} & $[-0.52,-0.49]$ & $[-0.06,0.04]$  & - \\
  \multicolumn{2}{c|}{\cite{Epelbaum:2012ua}} & $-1.9$ & $-0.05$ & - \\
  \multicolumn{2}{c|}{\cite{Cohen:1998jr}} & $-3.3$ & $-0.95$ & - \\
  \bottomrule[2pt]
  \label{tab:v2fit}
  \end{tabular}}
\end{table}

To further validate the method, we have also extracted the shape parameter $v_2$ and compared it with other studies. As shown in Table~\ref{tab:v2fit}, the results from both the latest NLO low-energy theorems (LETs)~\cite{Epelbaum:2015sha,Baru:2015ira} and the next-to-next-to-next-to-leading order LETs~\cite{Thim:2024jdv} are broadly consistent with the empirical values~\cite{deSwart:1995ui,PavonValderrama:2005ku,NavarroPerez:2014ovp}, whereas they differ significantly from the earlier leading-order LETs~\cite{Epelbaum:2012ua} and from methods that treat one-pion exchange perturbatively~\cite{Cohen:1998jr}. In the channel-by-channel fits, our extracted values for both the $|^1S_0,I=1\rangle$ and $|^3S_1,I=0\rangle$ channels agree well with the results of Refs.~\cite{deSwart:1995ui,PavonValderrama:2005ku,NavarroPerez:2014ovp,Epelbaum:2015sha,Baru:2015ira,Thim:2024jdv}. In the joint analysis, their $|^1S_0,I=1\rangle$ values mostly fall within the error margins of our extraction, while our $|^3S_1,I=0\rangle$ result remains close to them, even though we have not considered the $S$-$D$ mixing arising from the OPE tensor force. This consistency supports the applicability of our generalized ERE with the lhc, the $|^3S_1\rangle$ comparison being understood at the single-channel level ($S$-$D$ mixing neglected, as discussed above).

\section{Summary}\label{summarychapter}
We have applied the generalized ERE with the lhc~\cite{Du:2024snq} to low-energy $NN$ scattering, with a focus on the $np$ $^1S_0$ channel that exhibits an amplitude zero ($\delta=0$) near $k\simeq 0.35$~GeV (c.m. momentum), which cannot be accommodated using the traditional ERE without changes. Using the $np$ phase shifts of Ref.~\cite{Stoks:1993tb}, we performed fits at several expansion orders of Eq.~(\ref{equ:kcotdeltawithlhc}), including separate fits to the $\ket{^1 S_0, I=1}$ and $\ket{^3 S_1, I=0}$ channels and a joint fit constrained by a common $\pi N$ coupling constant and the coherent scattering length.\par

The extracted low-energy parameters, including the scattering lengths, effective ranges, shape parameter, and subthreshold poles (a virtual state in $^1 S_0$ and a bound state in $^3 S_1$), are consistent with established determinations and are stable across different expansion orders. Notably, the pseudoscalar $\pi N$ coupling constant $g_{\pi N}^2/(4\pi)$ is extracted within the OPE lhc approximation adopted here, relying primarily on analytic constraints from the lhc and scattering data. The results, shown in Table~\ref{tab:1S03S1fit}, are consistent with previous determinations from various independent methods, supporting the applicability of the framework in this setting. We find that the central values of $g_{\pi N}^2/(4\pi)$ extracted in the joint analysis tend to be smaller than several recent high-precision determinations around $g_{\pi N}^2/(4\pi)\simeq 14$, which may be understood as contributions beyond the present OPE lhc approximation (two-pion exchange, $S$-$D$ mixing, and $NN\pi$ and $N\Delta$ coupled channels); those are not included here. We therefore regard the extracted coupling as an effective measure of the OPE lhc strength, as well as the validity of our method, rather than a competitive high-precision determination.

These results indicate that explicitly including the lhc into the ERE not only improves the description of low-energy scattering observables but also provides a useful framework for constraining the long-range lhc contribution from phase-shift data. The generalized ERE with the lhc thus offers a systematic parametrization for near-threshold systems with light-particle exchanges, enhancing our understanding of the role of OPE in low-energy hadronic interactions.\par

\begin{acknowledgments} 

	We are grateful to Xiong-Hui Cao, Hao-Jie Jing, Bingwei Long, Qi-Fang Lü, and Mao-Jun Yan for useful discussions. This work is supported in part by the National Natural Science Foundation of China under Grants No.~12547115, No.~12125507, No.~12361141819, No.~12447101, and No.~12547111; by the National Key R\&D Program of China under Grant No.~2023YFA1606703; and by the Chinese Academy of Sciences under Grant No.~YSBR-101. MLD gratefully acknowledges the support of the Peng Huan-Wu Visiting Professorship and the hospitality of the Institute of Theoretical Physics, Chinese Academy of Sciences, where part of this work was completed. UGM also acknowledges the support of the CAS President's International Fellowship Initiative (PIFI) (Grant No. 2025PD00022), from the
    European Research Council (ERC) under the European Union's Horizon 2020 research and innovation programme (grant agreement No. 101018170) and 
    from Deutsche Forschungsgemeinschaft (DFG) under Grant No. 525056915
     under Germany's Excellence Strategy -- EXC 3107 -- Project-ID~533766364.

\end{acknowledgments}

\bibliography{refs}
\end{document}